\documentclass[journal,article,accept,moreauthors,pdftex,10pt,a4paper,sensors]{Definitions/mdpi} 
\firstpage{1} 
\pubvolume{xx}
\issuenum{1}
\articlenumber{1}
\pubyear{2018}
\copyrightyear{2018}
\externaleditor{Academic Editor: name}
\history{Received: date; Accepted: date; Published: date}

\Title{Towards End-to-End Acoustic Localization using Deep Learning:
  from Audio Signal to Source Position Coordinates}

\Author{Juan Manuel Vera-Diaz, Daniel Pizarro\orcidA{} and
  Javier Macias-Guarasa\orcidC{}$^*$}

\AuthorNames{Juan Manuel Vera-Diaz, Daniel Pizarro, Javier Macias-Guarasa}

\address[1]{%
  Department of Electronics, University of Alcalá, Campus Universitario
  s/n, 28805, Alcalá de Henares, Madrid, Spain. \quad
  manuel.vera@edu.uah.es, \quad daniel.pizarro@uah.es, \quad  javier.maciasguarasa@uah.es}

\corres{Correspondence: javier.maciasguarasa@uah.es; Tel.: +34-91-885-6918}

\abstract{ This paper presents a novel approach for indoor acoustic
  source localization using microphone arrays and based on a Convolutional
  Neural Network (CNN). The proposed solution is, to the best of our
  knowledge, the first published work in which the CNN is designed to
  directly estimate the three dimensional position of an
  acoustic source, using the raw audio signal as the input
  information avoiding the use of hand crafted audio features. 
  Given the limited amount of available localization data, we propose 
  in this paper a training strategy based on two steps. We first train our network using
  semi-synthetic data, generated from close talk speech recordings, and where we simulate the
  time delays and distortion suffered in the signal that propagates from the source to the array of microphones. 
  We then fine tune this network using a small amount of real data. 
  Our experimental results show that this strategy is able to produce networks that significantly improve existing 
  localization methods based on \textit{SRP-PHAT} strategies. In addition, our experiments show that our CNN  method exhibits better resistance
  against varying gender of the speaker and different window sizes compared with the other methods.}  

\newcommand{\legendMOTPDelta}{$\begin{array}{c}   MOTP(m) \\   \Delta_{r}^{MOTP} \end{array}$}

\newcommand{\seq}[1]{\texttt{s{#1}}}

\newcommand{\spm}{\;\;}
\newcommand{\mbf}{\mathbf}

\definecolor{greenao}{rgb}{0.0, 0.5, 0.0}

\newcommand{\lparen}{\left(}
\newcommand{\rparen}{\right)}

\newcommand*{\mathcolor}{}
\def\mathcolor#1#{\mathcoloraux{#1}}
\newcommand*{\mathcoloraux}[3]{%
  \protect\leavevmode
  \begingroup
    \color#1{#2}#3%
  \endgroup
}

\begin{document}


\section{Introduction}
\label{sec:introduction}


The development and scientific research in advanced perceptual systems
has notably grown during the last decades, and has experimented a
tremendous rise in the last years due to the availability of increasingly 
sophisticated sensors, the use of computing nodes with higher and
higher computational power, and the advent of powerful algorithmic
strategies based on deep learning (all of them actually entering the
mass consumer market). The aim of perceptual systems is to automatically
analyze complex and rich information taken from different sensors, in
order to obtain refined information on the sensed environment and the
activities being carried out within them. The scientific works in these
environments, cover research areas from basic sensor technologies, to
signal processing and pattern recognition, and open the path to the idea
of systems able to analyze human activities, providing them with
advanced interaction capabilities and services..


In this context, localization of humans (being the most
\emph{interesting} element for perceptual systems) is a fundamental task
that needs to be addressed so that the systems can actually start to
provide higher level information on the activities being carried
out. Without a precise localization, further advanced interaction
between humans and their physical environment cannot be carried out
successfully.


The scientific community has devoted a huge amount of effort to build
robust and reliable indoor localization systems, based on different
sensors~\cite{torres2010review,ruiz2010survey,mainetti2014survey}. 
Non-invasive technologies are preferred in this context, so that no
electronic or passive devices need to be carried by humans for
localization. The two non-invasive technologies
that have been mainly used in indoor localization are those based on
video systems and acoustic sensors.





This paper focuses on audio-based localization, with no previous
assumptions on the acoustic signal characteristics nor in the physical
environment, apart from the fact that unknown wide-band audio sources
(e.g. human voice) are captured by a set of microphone arrays placed in
known positions.
The main objective of the paper is to directly use the signals captured
by the microphone arrays to automatically obtain the position of the
the acoustic source detected in the given environment.

Even though there are a lot of proposals in this area, Acoustic Source Localization (ASL) is still a
hot research topic.  
This paper proposes a convolutional neural network (CNN) architecture that is trained end-to-end to solve the acoustic localization problem. To our knowledge, this is the first work in the literature that does not
provide the network with feature vectors extracted from the speech
signals, but directly uses the speech signal. Avoiding hand crafted features has been proved to increase the accuracy of classification and regression methods based on convolutional neural networks in other fields, such as in computer vision \cite{2012imagenet,2014vgg}.  

Our proposal is evaluated
using both semi-synthetic and real data, outperforming traditional solutions
based on Steered Response Power (\emph{SRP})~\cite{dibiase2000high},
that are still the basis of state-of-the-art
systems~\cite{nunes2014steered,cobos2017steered,he2018steered,salvati2018sensitivity}.

The rest of the paper is organized as follows. In
Section~\ref{sec:state-art} a review study of the state-of-the-art in
acoustic source localization with special emphasis on the use of deep
learning approaches. Section~\ref{sec:system-description} describes
the CNN based proposal, with details on the training and fine tuning
strategies. The experimental work is detailed in
Section~\ref{sec:experiments-and-discussion}, and
Section~\ref{sec:conclusions} summarizes the main conclusions and
contributions of the paper and gives some ideas for future
work.




\section{State of the Art}
\label{sec:state-art}

Many approaches exist in the literature to address the acoustic source
localization (ASL) problem. According to the classical literature review
in this topic, these approaches can be broadly divided in three
categories~\cite{Brandstein97practical,dibiase2001robust}: time delay
based, beamforming based, and high-resolution spectral-estimation based
methods. This taxonomy relies in the fact that ASL has been
traditionally considered a signal processing problem based on the
definition of a signal propagation
model~\cite{knapp1976GCC,Brandstein97practical,dibiase2001robust,Zhang2008PHAT,Dmochowski2010Steered,cobos2011modified,Butko2011,habets2010MVDR,marti2013},
but, more recently, the range of proposals in the literature also
considered strategies based on exploiting optimization techniques and
mathematical properties of related
measurements~\cite{velasco2012-F,padois2015,velasco2016denoising,compagnoni2017,salari2018},
and also using machine learning
strategies~\cite{murray2009,deleforge2013,salvati2016}, aimed at
obtaining a direct mapping from specific features to source
locations~\cite{rascon2017}, area in which deep learning approaches are
starting to be applied and that will be further described later in this
section.

Time delay based methods (also referred to as \emph{indirect methods}),
compute the time difference of arrivals (TDOAs) across various
combinations of pairs of spatially separated microphones, usually using
the Generalized Correlation Function (GCC)~\cite{knapp1976GCC}. In a
second step, the TDOAs are combined with knowledge of the microphones'
positions to generate a position
estimation~\cite{Brandstein97practical,stoica2006}.


Beamforming based
techniques~\cite{dibiase2001robust,Dmochowski2010Steered,marti2013,cobos2017} attempt to
estimate the position of the source, optimizing a spatial statistic
associated with each position, such as in the Steered Response Power
(\emph{SRP}) approach, in which the statistic is based on the signal
power received when the microphone array is steered in the direction of
a specific location. \emph{SRP-PHAT} is a widely used algorithm for
speaker localization based on beamforming that was first proposed
in~\cite{dibiase2000high}\footnote{Although the formulation is virtually
  identical to the \emph{Global Coherence Field} (GCF) described in
  ~\cite{omologo1993}}. It combines the robustness of the SRP approach
with the Phase Transform (PHAT) filtering, which increases the
robustness of the algorithm to signal and room conditions, making it an
ideal strategy for realistic speaker localization
systems~\cite{Dmochowski2007,Badali2009,Hoang2010,cobos2011modified,Butko2011}. Other
beamforming based methods such as the Minimum Variance Distortionless
Response (MVDR)~\cite{habets2010MVDR}, exhibits problems when facing
reverberant environments, because it introduces a new trade-off between
dereverberation and noise reduction.

In what respect to spectral estimation based methods, the multiple
signal classification algorithm (MUSIC)~\cite{Schmidt1986}, has been
widely used, but these methods, in general, tend to be less robust than
beamforming methods~\cite{dibiase2001robust}, as they assume incoherent
signals and are very sensitive to small modeling errors.




In the past few years, deep learning
approaches~\cite{goodfellow2016deep} have taken the lead in different
signal processing and machine learning fields, such as computer
vision~\cite{krizhevsky2012imagenet,He2016DeepRL} and speech
recognition~\cite{hinton2012deep,graves2014towards,Deng2014EnsembleDL}, and, in general, in
any area in which complex relationships between observed signals and the
underlying processes generating them need to be discovered.


The idea of using neural networks for ASL is not new. Back in the early
nineties and the first decade of the current century, works such
as~\cite{steinberg1991,datum1996,murray2009} proposed the use of neural
network techniques in this area. However an evaluation on realistic and
extensive data sets was not viable at this time, and the proposals were
somehow limited in scope.

With the advent and huge increase on applications of deep neural
networks in all areas of machine learning, and mainly due to the
sophisticated capabilities and more careful implementation details of
network architectures and the availability of advanced hardware
architectures with increased computational capacity, promising works
have been proposed also for ASL~\cite{youseff2013RobustBinauralSL,Xiao2015ALA,ma2015ExploitDNNbinauralloc,Takeda2016DiscriminativeMS,Takeda2016SoundSL,takeda2017unsupervisedadaptationDNN,sun2018ProbabNN,Chakrabarty2017b,Yalta2017ASLDeepLearningModels,ferguson2017CNNs,hirvonen2015,he2017DNNMultipleSpkLocal,Salvati2018CNNsImprovedASL,Ma2018PhasedMicArray,thuillier2018AudioFeatDiscovery}.


The main differences between the different proposals using neural
networks for ASL reside in the architectures, input features, the network
output (target), and the experimental setup (using real or simulated
data).

Regarding the information given to the neural network, we can find
several works using features physically related to the ASL problem. Some
of the proposals use features derived from the GCC or related functions,
which actually make sense as these correlation function is closely
related to the TDOAs which are used in traditional methods to generate
position estimations. The published works use either the GCC coefficients
directly~\cite{sun2018ProbabNN}, features derived from
them~\cite{Xiao2015ALA,he2017DNNMultipleSpkLocal} or from the
correlation matrix~\cite{Takeda2016DiscriminativeMS,takeda2017unsupervisedadaptationDNN}, or even combined
with others, such as cepstral coefficients~\cite{ferguson2017CNNs}. Other
works are focused in exploiting binaural
cues~\cite{youseff2013RobustBinauralSL,ma2015ExploitDNNbinauralloc},
features derived from convolving the spectrum with head related impulse
responses~\cite{thuillier2018AudioFeatDiscovery} or even narrowband SRP
values~\cite{Salvati2018CNNsImprovedASL}. The latter approach goes one
step further from correlation related values, as the SRP function
actually integrates multiple GCC estimations in such a way that acoustic
energy maps can be easily generated from it.

Opposed to the previously described works using refined features directly
related to the localization problem, we can also find others using
frequency domain features
directly~\cite{Takeda2016SoundSL,Yalta2017ASLDeepLearningModels}, in
some cases generated from spectrograms of general time-frequency
representations~\cite{Chakrabarty2017b,hirvonen2015}. These approaches
represent a step forward compared with the previous ones, as they give
the network the responsibility of automatically learn the relationship
between spectral cues and the location related
information \cite{Ma2018PhasedMicArray} kind of combines both
strategies, as they use spectral features but calculating them in a
cross-spectral fashion, that is, combining the values from all the
available microphones in the so-called Cross Spectral Map (CSM).

In none of the referenced works, the authors try to make use of the raw
acoustic signal directly, and we are interested in evaluating the
capabilities of CNN architectures in directly exploiting this raw input
information.

In what respect to the estimation target, most of the works are oriented
towards estimating the Direction of Arrival (DOA) of the acoustic
sources~\cite{Xiao2015ALA,sun2018ProbabNN,Chakrabarty2017b,he2017DNNMultipleSpkLocal,Salvati2018CNNsImprovedASL},
or DOA related measurements such as azimuth
angle~\cite{youseff2013RobustBinauralSL,ma2015ExploitDNNbinauralloc,Takeda2016SoundSL},
elevation angle~\cite{thuillier2018AudioFeatDiscovery}, or position
bearing+range~\cite{ferguson2017CNNs}. Some of the proposals pose the
problem not as a direct estimation (regression) but as a classification
problem among a predefined set of possible position related
values~\cite{hirvonen2015,Takeda2016SoundSL,Takeda2016DiscriminativeMS,takeda2017unsupervisedadaptationDNN,Yalta2017ASLDeepLearningModels}
(azimuth, positions in a predefined grid, etc.). Works with a very
different target try to estimate a \emph{clean} acoustic source
map~\cite{Ma2018PhasedMicArray} or learn time-frequency masks as a
preprocessing stage prior to ASL~\cite{pertila2017}.

In none of the referenced works the authors try to directly estimate the
coordinate values of the acoustic sources, and, again, we are interested
in evaluating the capabilities of CNN architectures to directly
generate this output information.

Finally, in what respect to the experimental setup, most works use
simulated data either for training or for training and
testing~\cite{youseff2013RobustBinauralSL,Xiao2015ALA,ma2015ExploitDNNbinauralloc,Takeda2016DiscriminativeMS,Takeda2016SoundSL,takeda2017unsupervisedadaptationDNN,sun2018ProbabNN,Chakrabarty2017b,pertila2017,Yalta2017ASLDeepLearningModels,hirvonen2015,he2017DNNMultipleSpkLocal,Salvati2018CNNsImprovedASL,Ma2018PhasedMicArray,thuillier2018AudioFeatDiscovery},
usually by convolving clean (anechoic) speech with impulse responses
(room, head related, or DOA related (azimuth, elevation)). Only some of
them actually face real
recordings~\cite{youseff2013RobustBinauralSL,Xiao2015ALA,ferguson2017CNNs,he2017DNNMultipleSpkLocal,Salvati2018CNNsImprovedASL},
which in our opinion is a must to be able to assess the actual impact of
the proposals in real conditions.

So, in this paper we describe, for the first time in the literature to
the best of our knowledge, a CNN architecture in which we directly
exploit the raw acoustic signal to be provided to the neural network,
with the objective of directly estimating the three dimensional position
of an acoustic source in a given environment. This is the reason why we
refer to this strategy as end-to-end, considering the full coverage of
the ASL problem. The proposal has been tested on both semi-synthetic and real
data from a publicly available database.

\section{System Description}
\label{sec:system-description}


\subsection{Problem Statement}
\label{sec:problem-statement}

Our system obtains the position of an acoustic source from the audio signals recorded 
by an array of $M$ microphones. Given a reference coordinate origin, the source position 
is defined with the 3D coordinate vector $\mbf{s}=\lparen{} s_x \spm s_y \spm s_z \rparen^\top$. 
The microphones positions are known and they are defined with coordinate vectors $\mbf{m}_i=\lparen{}m_{i,x} \spm m_{i,y} \spm m_{i,z}\rparen^\top$ with $i=1,\dots,M$. 
The audio signal captured from the $i^{th}$ microphone is denoted by $x_i(t)$. This signal 
is discretized with a sampling frequency $f_s$ and is defined with $x_i[n]$. We assume 
for simplicity that $x_i[n]$ is of finite-length with $N$ samples. This
corresponds to a small window of audio with duration $w_s=N/f_s$, which
is a design parameter in our system. We denote as $\mbf{x}_i$ the vector
containing all time samples of the signal: 

\begin{equation}
  \label{eq:vector}
  \mbf{x}_i = \begin{pmatrix}x_i[0]&\dots&x_i[N-1]\end{pmatrix}^\top. 
\end{equation}

The problem we seek to solve is to find the following regression function $f$: 
\begin{equation} \label{eq:eqsystem}
  \mbf{s} = f\lparen\mbf{x}_1, \dots , \mbf{x}_{N}, \mbf{m}_1, \dots, \mbf{m}_M \rparen,  
\end{equation}
that obtains the speaker position given the signals recorded from the microphones.


In classical simplified approaches, $f$ is found by assuming that signals received from different microphones mainly differ by a delay that depends on the relative position of the source with respect to the microphones.  However, this assumption breaks in environments where the signal suffers from random noise and distortion, such as multi-path signals or microphone non-linear response. 

Due to the aforementioned effects, and the random nature of the audio signal, the regression function of equation (\ref{eq:eqsystem}) cannot be estimated analytically. We present in this paper a learning approach for directly obtaining $f$ using Deep Learning. We represent $f$ using a Convolutional Neural Network (CNN) which is learned end-to-end from the microphone signals. In our system we assume that microphones positions are fixed. We thus drop the requirement of knowing the microphone's position from equation (\ref{eq:eqsystem}) which will be implicitly learned by our network with the following regression problem:

\begin{equation} \label{eq:eqsystem2}
  \mbf{s} = f_{net}(\mbf{x}_1, \dots , \mbf{x}_{M}),  
\end{equation}
where $f_{net}$ denotes the function that we represent using the CNN and
whose topology is described next.

\subsection{Network Topology}
\label{sec:network-topology}

\begin{figure}[H]
	\centering
	\includegraphics[width=0.9\textwidth]{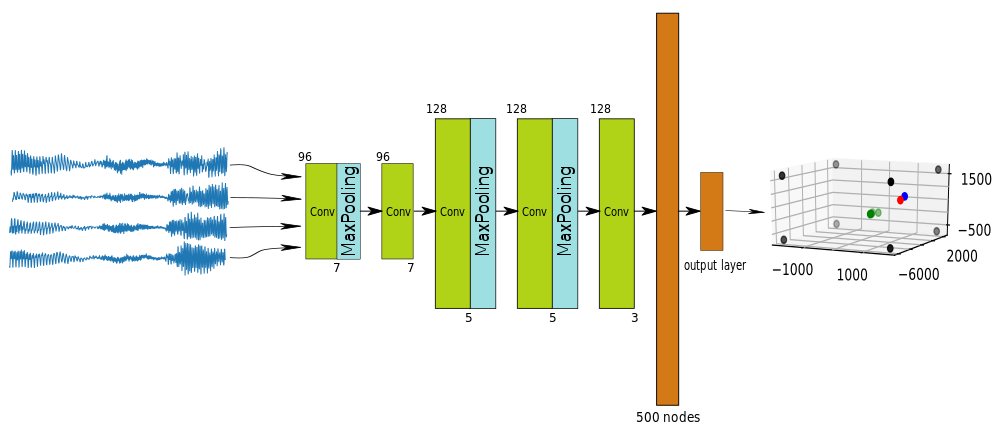}
	\caption{Used network topology}
	\label{fig:NetScheme}
\end{figure}

The topology of our neural network is shown in figure
\ref{fig:NetScheme}. It is composed of five convolutional blocks of one
dimension and two fully connected blocks. Following equation
(\ref{eq:eqsystem2}), the network inputs are the set of windowed signals
from the microphones and the network output is the estimated position of
the acoustic source.

Table (\ref{tab:NetworkLayers}) shows the size and amount of convolutional filters in the proposed network. We use filters of size $7$ (layers $1$ and $2$), size $5$ (layers $3$ and $4$) and size $3$ (layer $5$). The number of filters is $96$ in the first two convolutional layers and $128$ in the rest. As seen in figure \ref{fig:NetScheme}, some of the layers are equipped with \textit{MaxPooling} filters with the same pool size as their corresponding convolutional filters. The last two layers are fully-connected layers, one hidden with $500$ nodes and the output layer. All layer's activation functions are ``ReLUs''  with the exception of the output layer. During training we include dropout with probability $0.5$ in the fully-connected layers to prevent overfitting.   


\begin{table}[H]
	\centering
	\begin{tabular}{ccc}
    \hline
    \textbf{Block} & \textbf{Filters} & \textbf{Kernel}\\\hline
    Convolutional block 1 & 96 & 7 \\\hline
    Convolutional block 2 & 96 & 7 \\\hline
    Convolutional block 3 & 128 & 5 \\\hline
    Convolutional block 4 & 128 & 5 \\\hline
    Convolutional block 5 & 128 & 3 \\\hline
  \end{tabular}
  \caption{Network convolutional layers summary}
  \label{tab:NetworkLayers}
\end{table}

\subsection{Training Strategy}
\label{sec:training-strategy}

The amount of available real data that we have in our experimental setup
(see Section~\ref{sec:experiments-and-discussion}) will be, in general, limited for training a CNN model. 
To cope with this problem we propose a training strategy comprising two steps:

\begin{enumerate}
\item[] Step 1. Training the network with semi-synthetic data: 
  We use close-talk speech recordings and a set of randomly generated source positions
  to generate simulated versions of the signals captured by a set of microphones that share the same geometry with the environment used in real data. 
  Additional considerations on the acoustic behavior of
  the target environment (specific noise types, noise levels, etc.) 
  is also taken into account to generate the data. This dataset can virtually be made as big as required to train the network. 
  
\item[] Step 2.  Fine tuning the network with real data:  We train
  the network on a reduced subset of the database captured in the target physical
  environment using the weights obtained in Step 1 as initialization.
 
\end{enumerate}

\subsubsection{Semi-Synthetic Dataset Generation}
\label{sec:semi-synth}
  
In this step we extract audio signals from any available close-talk
(anechoic) corpus, and use them to generate semi-synthetic data. 
There are many available datasets suitable for this task (freely of
commercially distributed). Our semi-synthetic dataset can thus
 be made as big as required for training the CNN.


For this task, we randomly generate position vectors
$\mbf{q}=\lparen q_x \spm q_y \spm q_z \rparen^\top$ of the acoustic
source using a uniform distribution that covers the physical space
(room) that will be used.


The loss function we use to train the network is the mean squared error between the
estimated position given by the network ($\mbf{s_{i}}$) and the target position vector ($\mbf{q_{i}}$). It follows the expression:

\begin{equation} \label{eq:lossFunc}
  \mathcal{L}( \mbf{\Theta}) = \frac{1}{N} \displaystyle\sum_{i=1}^{N}\left|\mbf{q_{i}} - \mbf{s_{i}} \right|^2,
\end{equation}

where $\mbf{\Theta}$ represents the weights of the network. Equation (\ref{eq:lossFunc}) is minimized in function of the unknown weights using iterative optimization based on the Stochastic Gradient Descent (SGD) algorithm~\cite{le2011}. We finally obtain the target weights $\theta \in \mbf{\Theta}$ once a termination criterion is met in the optimization. More details are given in Section~\ref{sec:experiments-and-discussion} about the training algorithm. 


In order to realistically simulate the signals received in the microphones from a given
source position we have to consider two main issues:

\begin{itemize}
\item Signal propagation considerations: This is affected by the impulse
  response of the target room. Different alternatives can be used to
  simulate this effect, such as convolving the anechoic signals with
  real room impulse responses such as
  in~\cite{Takeda2016DiscriminativeMS}, that can be difficult to acquire
  for general positions in big environments; or using room response
  simulation methods such as the image
  method~\cite{allen1979imagemethod} used in~\cite{velasco2014-F} for
  this purpose.
\item Acoustic noise conditions of the room and recording process
  conditions: These can be due to additional equipment (computers, fans,
  air conditioning systems, etc.)  present in the room, and to problems
  in the signal acquisition setup. This can be addressed by assuming
  additive noise conditions, and selecting a noise type and acoustic
  effects that should be preferably estimated in the target room.
\end{itemize}

In our case, and regarding the first issue, we used an initial simple
approach, just taking into account the propagation delay from the source
position to each of the microphones, that depends on their relative
position and the sound speed in the room.


We denote the number of samples  we have to shift a signal to simulate the arrival delay suffered at microphone $i$ by $N_{s_{i}}=f_{s}\frac{d_{i}}{c}$ where $f_{s}$ is the sampling
frequency of the signal, $d_{i}$ is the euclidean distance between the acoustic source and the $i$ microphone and $c$ is the sound speed in air ($c=343 m/s$ in a room at $20 Cº$). In general $N_{s_i}$ is not an integer number. We thus require a way to simulate sub-sample shifts in the signal. In order to implement the delay $N_{s_i}$  on $\mbf{x}_{pc}$ (the windowed signal of $N$ samples from the close-talk  dataset) to obtain $\mbf{x}_i$ we use the following transformation:

\begin{equation} \label{eq:eqFFT}
  \mbf{X}_{pc} = \mathcal{F} \lbrace \mbf{x}_{pc} \rbrace \spm\spm  \mbf{x}_i =A_i\lparen\mathcal{F}^{-1} \lbrace \mbf{X}_{pc}\, \mbf{D}_{s_i} \rparen,\;\;\;with\;\mbf{D}_{s_i}\nobreak = \nobreak\left( 1, e^{-j \frac{2 \pi N_{s_i}}{N}}, e^{-j \frac{4 \pi N_{s_i}}{N}}, \cdots , e^{-j (N-1) \frac{2 \pi N_{s_i}}{N}} \right)
\end{equation}
where we first transform $\mbf{x}_{pc}$ into the frequency domain $\mbf{X}_{pc}$ using the \textit{Discrete Fourier Transform} operator $\mathcal{F}$. We then change its phase according to $N_{s_i}$ by the phase vector $\mbf{D}_{s_i}$ and transform the signal back into time domain $\mbf{x}_i$, using the \textit{Inverse Discrete Fourier Transform} operator $\mathcal{F}^{\emph{\tiny{-1}}}$. $A_i$ is an amplitude factor applied to the signal that follows a uniform random distribution, and it is different for each microphone, preventing the network from being affected by amplitude differences between the signals captured in different microphones ($A_i \in [0.01, 0.03]$ in the experimental setup described in Section~\ref{sec:experiments-and-discussion}).




Regarding the second issue, we simulate noise and disturbances in the
signals arriving to the microphones so that the signal-to-noise ratio
and the spectral content of the signals are as similar as possible to those found in the real data.    
In order to  provide an example of the methodology we follow,
 we refer in this section to the particular case of the IDIAP room (see
Section~\ref{sec:idiap-av16.3-corpus}) that will be used in our real
data experiments, and the Albayzin Phonetic Corpus (see
Section~\ref{sec:albayz-phon-corp}) that will be used for synthetic data
generation.

In the IDIAP room, a spectrogram based analysis showed that the
recordings are contaminated with a tone at around $25Hz$ in the spectrum
which does not appear in anechoic conditions, probably due to room
equipment of electrical noise generated in the recording hardware
setup. We have determined that the frequency of this tone actually
varies in a range between $20Hz$ and $30Hz$. So, in the synthetic data
generation process, we have \emph{contaminated} the signals from the
phonetic corpus with an additive tone of a random frequency in this
established range, and we have also added white gaussian noise following
the expression:

\begin{equation} \label{eq:signalAdded}
  x_{pc_{new}}[n] =  x_{pc}[n] + k_s\sin (2\pi f_{0} n/ f_{s} +  \phi_{0}) + k_\eta \eta_{wgn}[n],
\end{equation}
where $k_s$ is a scaling factor for the contaminating tone signal
(similar to the tone amplitude found in the target room recordings,
$0.1$ in our case), $f_{0} \in [20, 30]Hz$, $\phi_0 \in [0, \pi]
rad$, $\eta_{wgn}$ is a white gaussian noise signal, and $k_\eta$ is a
noise scaling factor to generate signals with a SNR which is similar to
that found in the target room recordings.

After this procedure is applied, the semi-synthetic signal data set will
be ready to be used in the neural network training procedure.


\subsubsection{Fine Tuning Procedure}
\label{sec:fine-tuning}

The previous step takes care of reproducing simple acoustic
characteristics of the testing room 
such as the propagation effects and the presence of specific types and
levels of additive noises, but there are other phenomena like multi-path
and reverberation propagation which are more complex to simulate. In
order to introduce these acoustic behaviors of the target physical
environment, our proposal is to carry out a fine tuning procedure of the
network model using a short amount of real recorded data in the target
room

Although there are other methods such as the one proposed in
\cite{takeda2017unsupervisedadaptationDNN}, where an unsupervised DNN is
implemented for the adaptation of parameters to unknown data, we believe
that the fine tuning process implemented is adequate because, in the
first place, it is a supervised process with which a better performance
is expected to be obtained and, secondly, not all the sequences of the
test data set are used, so that only a few are used for the fine tuning
process, saving the rest for the test phase.








\section{Experimental Work}
\label{sec:experiments-and-discussion}
    
In his section we describe the datasets used in both steps of the
training strategy described in Section~\ref{sec:training-strategy}, and
the details associated with it.  We then define the experimental setup
general conditions, and the error metrics used for comparing our
proposal with other state-of-the-art methods and finally present our
experimental results, starting from the baseline performance we aim at
improving.

\subsection{Datasets}
\label{sec:datasets}

\subsubsection{IDIAP AV16.3 Corpus: for testing and fine tuning}
\label{sec:idiap-av16.3-corpus}

We have evaluated our proposal using the audio recordings of the AV16.3
database~\cite{lathoud2005av16}, an audio-visual corpus recorded in the
\emph{Smart Meeting Room} of the IDIAP research institute, in
Switzerland. We have also used the physical layout of this room for our
semi-synthetic data generation process.

\begin{figure}[H]
	\begin{tabular}{ccc}
		\includegraphics[width=0.35\textwidth]{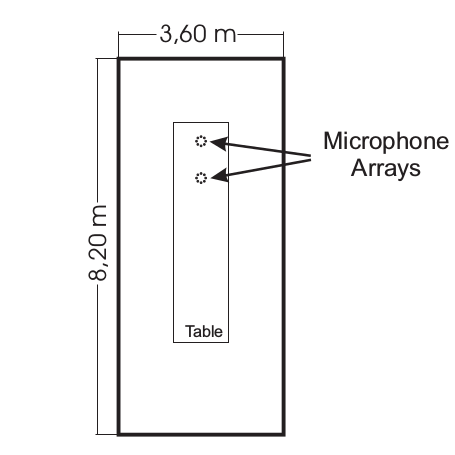}
		& \includegraphics[width=0.4\textwidth]{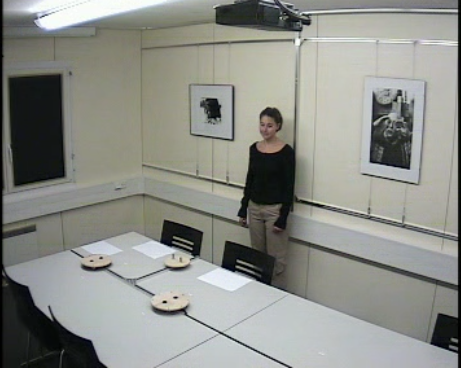}
		& \includegraphics[width=0.175\textwidth]{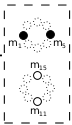}\\
		\textbf{(a)} & \textbf{(b)} & \textbf{(c)}
	\end{tabular}
	\caption{(a) Simplified top view of the \textit{IDIAP Smart Meeting
      Room}, (b) A real picture of the room extracted from a video
    frame, (c) Microphone setup used in this proposal}
	\label{fig:IdiapSettings}
\end{figure}

The \textit{IDIAP} \textit{Smart Meeting Room} is a $3.6m \times 8.2m
\times 2.4m$ rectangular room with a rectangular table centrally located
and measuring $4.8m \times 1.2m$. On the table's surface there are two circular
microphone arrays of $0.1m$ radius, each of them composed by 8 regularly
distributed microphones as shown in figure \ref{fig:IdiapSettings}. The
centers of both arrays are separated by a distance of $0.8m$. The middle point
between them is considered as the origin of the coordinate reference
system. A detailed description of the meeting room can be found in
~\cite{moore2002}.

The dataset is composed by several sequences of recordings, synchronously sampled at 16 KHz,
which a wide range of experimental conditions in the number of speakers
involved and their activity. Some of the available audio sequences are assigned a
corresponding annotation file containing the real ground truth positions
(3D coordinates) of the speaker's mouth at every time frame in which
that speaker was talking. The segmentation of acoustic frames with
speech activity was first checked manually at certain time instances by
a human operator in order to ensure its correctness, and later extended
to cover the rest of recording time by means of interpolation
techniques. The frame shift resolution was defined to \linebreak be 40
ms.  The complete dataset is fully accessible on-line at \cite{av163}.

In this paper we will just focus on all the annotated sequences of this
dataset featuring a single speaker, whose main characteristics are shown
in Table \ref{tab:DataInfo}. This allows us to directly compare our
performance with the state-of-the-art method presented
in~\cite{velasco2012-F}. Note that the firsts three sequences are
performed by a speaker remaining static while speaking at different
positions, and the last two ones by a moving speaker, being all of the
speakers different. We will refer to these sequences as \seq01, \seq02,
\seq03, \seq11 and \seq15 for brevity.

\begin{table}[H]
	\centering
	\begin{tabular}{c p{0.1\textwidth}  p{0.1\textwidth}  p{0.1\textwidth}  p{0.45\textwidth}}
		\hline
		\textbf{Sequence}	& \textbf{Average speaker height (cm)$^*$} & \textbf{Duration (seconds)}	& \textbf{Number of
      ground truth frames}	& \textbf{Description}\\
		\hline		\hline
		seq01-1p-0000 & 54.3 & 208 & 2248 & A single male speaker, static while speaking, at each of 16 locations. The speaker is facing the microphone
    arrays.\\\hline
		seq02-1p-0000 & 62.5 & 171 & 2411 & A single female speaker, static while speaking, at each of 16 locations. The speaker is facing the microphone
    arrays.\\\hline
		seq03-1p-0000 & 70.3 & 220 & 2636 &  A single male speaker, static while speaking, at each of 16 locations. The speaker is facing the microphone
    arrays.\\\hline
		seq11-1p-0100 & 53.5 & 33 & 481 & A single male speaker, making
                                      random movements while speaking,
                                      and facing the arrays. \\\hline
		seq15-1p-0100 & 79.5 & 36 & 436 & A single male speaker, walking around while alternating speech and long silences. No constraints\\\hline
	\end{tabular}
  \footnotesize{$^*$ The average speaker height is referenced to the system
  coordinates and refers to the speaker's mouth height.}
	\caption{\textit{IDIAP Smart Meeting Room} used sequences.}
	\label{tab:DataInfo}
\end{table}



\subsubsection{Albayzin Phonetic Corpus: for Semi-Synthetic Dataset Generation}
\label{sec:albayz-phon-corp}

The Albayzin Phonetic Corpus~\cite{albayzinELRA} consists of 3 sub-corpora
of 16 kHz 16 bits signals, recorded by 304 Castilian Spanish speakers
in a professional recording studio using high quality
close talk microphones.

We use this dataset to generate semi-synthetic data as described 
in Section~\ref{sec:semi-synth}. From the 3 sub-corpora, we will be only using the so-called
\emph{phonetic corpus}~\cite{MorenoAlbayzin1993}, composed of 6800
utterances of phonetically balanced sentences. This phonetical balance
characteristic makes this dataset perfect for generating our semi-synthetic data, 
as it will cover all possible acoustic contexts.

\subsection{Training and Fine Tuning Details}
\label{sec:training-details}

In the semi-synthetic dataset generation procedure, described in
Section~\ref{sec:semi-synth}, we generate random positions $\mbf{q}$
with uniformly distributed values in the following intervals:
$q_x \in [0, 3.6] m$, $q_y \in [0, 8.2] m$ and $q_z \in [0.92, 1.53]m$,
which correspond to the possible distribution of the speaker's mouth positions in
the \textit{IDIAP} room~\cite{lathoud2005av16}.

Regarding the optimization strategy for the loss function
described by equation~(\ref{eq:lossFunc}) we employ the
 \textit{ADAM}~\cite{adam2014} optimizer (variant of SGD with variable learning rate) 
along 200 epochs with a batch size of 100 samples. 
7200 different frames of input data per epoch are
randomly generated during the training phase and other 800 for validation.

The experiments will be performed with three different window lengths
($80 ms$, $160 ms$ and $320 ms$), so the training phase will be run once
per window length, obtaining three different network models. In each
training, 200 audio recordings are randomly chosen and 40 different
windows are randomly extracted from each. In the same way, 200 acoustic
source position $\mbf{q}$ vectors are randomly generated so that each
position generates 40 windows of the same signal. 

For the fine tuning procedure described in
Section~\ref{sec:fine-tuning}, we will be mainly using sequences \seq11
and \seq15, that features a speaker moving in the room while speaking,
and also sequences \seq01, \seq02 and \seq03 in a final
experiment.  


As it will be described in Section~\ref{sec:results-discussion}, we will
also address experiments trying to assess the relevance of adding
additional sequences \seq01, \seq02 and \seq03 to complement the fine
tuning data provided by \seq11 and \seq15.  We will also refer to gender
and height issues in the fine tuning and evaluation data.



\subsection{Experimental Setup}
\label{sec:experimental-setup}

In our experiments, sequences \seq01, \seq02 and \seq03  are used
 for testing the performance of our network and, as explained above, to complement sequences \seq11 and \seq15 for fine tuning.  

 In this work, we are using a simple microphone array configuration,
 aimed at evaluating our proposal in a resource-restricted environment,
 as it was done in~\cite{velasco2012-F}. In order to do so, we are using
 4 microphones (numbers 1, 5, 11 and 15, out of the 16 available in the
 AV16.3 data set), grouped in two microphone pairs. The selected
 microphone pairs configurations are shown in
 Figure~\ref{fig:IdiapSettings}.c, in which microphones with the same
 color are considered as belonging to the same microphone pair.  We
 provide results depending on the length of the acoustic frame, for $80 ms$,
 $160 ms$ and $320 ms$, to precisely assess to what extent the improvements are
 consistent with varying acoustic time resolutions.

 The main interest of our experimental work is assessing whether the
 end-to-end CNN based approach (that we will refer to as CNN) is
 competitive as compared with state-of-the-art localization methods.  We
 will compare this CNN approach with the standard \textit{SRP-PHAT}
 method, and the recent strategy proposed in~\cite{velasco2012-F} that
 we will refer to as GMBF.  This GMBF method is based on fitting a
 generative model to the GCC-PHAT signals using sparse constraints, and
 it reported significant improvements over \textit{SRP-PHAT} in the
 \textit{IDIAP} dataset~\cite{velasco2012-F,JoseVelascoThesis2017}.





 After providing baseline results comparing \textit{SRP-PHAT}, GMBF and
 our proposal without fine tuning procedure, we will then describe four
 experiments, that we briefly summarize here:

\begin{itemize}
\item In the first experiment, we will evaluate the performance
  improvements when using a single sequence for the fine tuning
  procedure.
\item In the second experiment, we will evaluate the differences between
  the semi-synthetic training plus the fine tuning approach, versus
  just training the network from scratch.
\item In the third experiment, we will evaluate the impact of adding an
  additional fine tuning sequence.
\item In the last experiment, we will evaluate the final performance
  improvements when also adding static sequences to the refinement
  process.
\end{itemize}

\subsection{Evaluation metrics}
\label{sec:evaluation-metrics}

Our CNN based approach yields a set of spatial coordinates
$\mbf{s}_k=\lparen{} s_{k,x} \spm s_{k,y} \spm s_{k,z} \rparen^\top$ that are
estimations of the current speaker position as time instant $k$. These position estimates
will be compared, by means of the Euclidean distance, to the ones
labeled in a transcription file containing the real positions
$\mbf{s}_{k_{GT}}$ (\emph{ground truth}), of the speaker.


We evaluate performance  adopting the same metric used in
\cite{velasco2012-F} and developed under the CHIL
project~\cite{mostefa2006clear}. It is known as MOTP (\textit{Multiple Object
  Tracking Precision}) and is defined as:

\begin{equation}
  MOTP=\frac{\displaystyle \sum_{k=1}^{N_{P}}|\mbf{s}_{k_{\tiny{GT}}}-\mbf{s}_{k}|^2}{N_{P}},
\end{equation}
where $N_{P}$ denotes the total number of position estimations along
time, $\mbf{s}_k$ the estimated position vector and $\mbf{s}_{k_{GT}}$
the labeled ground truth position vector. 

We will compare our experimental results, and that of the GMBF method,
with that of \textit{SRP-PHAT}, measuring the relative improvement in
MOTP with method, that is defined as follows:

\begin{equation}
  \Delta_{r}^{MOTP} = 100
  \frac{MOTP_{SRP-PHAT}-MOTP_{proposal}}{MOTP_{SRP-PHAT}} [\%]
\end{equation}

\subsection{Baseline Results}
\label{sec:baseline-results}

The baseline results are shown in Table \ref{tab:baselineResults} for
sequences \seq01, \seq02 and \seq03, and all the
evaluated time window sizes (in all the tables showing results in this
paper, \textbf{bold font} highlight the
best ones for a given data sequence and window length). The Table shows the results achieved by the
\textit{SRP-PHAT} standard algorithm strategy (columns SRP), the
alternative described in \cite{velasco2012-F} (columns GMBF), and the
proposal in this paper without applying the fine-tuning procedure
(columns CNN). We also show the relative improvements of GMBF and
CNN as compared with SRP-PHAT. 

\begin{table}[H]
\footnotesize
	\centering
	\begin{tabular}{c|ccc|ccc|ccc}
		      &   \multicolumn{3}{c|}{\textbf{80$ms$}}	& \multicolumn{3}{c|}{\textbf{160$ms$}} & \multicolumn{3}{c}{\textbf{320$ms$}}\\
   		    &   SRP & GMBF & CNN & SRP & GMBF & CNN & SRP & GMBF & CNN \\\hline\hline
    \multirow{2}{*}{\seq01 \legendMOTPDelta} &             $1.020$ &    $\mbf{0.795}$ &  $1.615$     & $0.910$ &    $\textbf{0.686}$ &     $1.526$ & $0.830$ &    $\mbf{0.588}$ &     $1.464$ \\
          &    & $\mbf{22.1\%}$ &  $-58.3\%$ &       & $\mbf{24.6\%}$ & $-67.7\%$ &       & $\mbf{29.1\%}$ & $-76.4\%$ \\\hline
    \multirow{2}{*}{\seq02 \legendMOTPDelta} &             $0.960$ & $\mbf{0.864}$    &      $2.124$ & $0.840$ & $\mbf{0.759}$    &     $1.508$ & $0.770$ & $\mbf{0.694}$    & $1.318$     \\
          &    & $\mbf{10.0\%}$ & $-121.3\%$ &       & $\mbf{9.6\%}$  & $-79.5\%$ &       & $\mbf{9.9\%}$  & $-71.2\%$ \\\hline
    \multirow{2}{*}{\seq03 \legendMOTPDelta} &  $0.900$ & $\mbf{0.686}$ & $1.559$ & $0.770$ & $\mbf{0.563}$ & $1.419$ & $0.690$ & $\mbf{0.484}$ & $1.379$\\
          &    & $\mbf{23.8\%}$ & $-73.2\%$ &  & $\mbf{26.9\%}$ & $-84.3\%$ &  & $\mbf{29.9\%}$ & $-99.9\%$\\\hline\hline
\multirow{2}{*}{Average \legendMOTPDelta}  & $0.957$ & $\mbf{0.778}$ & $1.763$ & $0.836$ & $\mbf{0.666}$ & $1.481$ & $0.760$ & $\mbf{0.585}$ & $1.385$\\
 &  & $\mbf{18.7\%}$ & $-84.3\%$ &  & $\mbf{20.4\%}$ & $-77.1\%$ &  & $\mbf{22.9\%}$ & $-82.3\%$\\
	\end{tabular}
	\caption{Baseline results for the SRP-PHAT strategy
    (columns SRP); the one in \cite{velasco2012-F} (columns
    GMBF), and the CNN trained with synthetic data without applying the
    fine-tuning procedure (columns CNN) for sequences \seq01,
    \seq02 and \seq03 for different window sizes. Relative
    improvements as compared to SRP-PHAT are shown below the MOTP
    values.}
	\label{tab:baselineResults}
\end{table}





The main conclusions from the baseline results are:

\begin{itemize}
\item 
  Best MOTP values for the
  standard SRP-PHAT algorithm are around $69cm$, with averages between $76cm$
  and $96cm$. For the GMBF, best MOTP values are around $48cm$, with
  averages between $59cm$ and $78cm$.
\item MOTP values improve as the frame size increases, as expected,
  given that better correlation values will be estimated for longer
  window signal lengths.
\item The GMBF strategy, as described in \cite{velasco2012-F},
  achieves very relevant improvements as compared with SRP-PHAT, with
  average relative improvements around $20\%$, and peak values of almost
  $30\%$.
\item Our CNN strategy, which at this point is only trained with semi-synthetic data,
    is very far from reaching the SRP-PHAT or GMBF in terms of performance. This result
    leads us to think that there are other effects only present in real data, such as reverberation, that are affecting
    the network.
\end{itemize}

Given the discussion above, we decided to apply the fine tuning strategy
discussed in Section~\ref{sec:fine-tuning}, with the experimental
details described in Section~\ref{sec:training-details}. So, the results
shown in Table~\ref{tab:baselineResults} will be compared with those
obtained by our CNN method, under different fine tuning (and training)
conditions, and will be described below.


\subsection{Results and Discussion}
\label{sec:results-discussion}

The first experiment in which we applied the fine tuning procedure used
\seq15 as the fine tuning subset. 


Table~\ref{tab:baselineResults+ft15} shows the results obtained by GMBF
(columns GMBF) and CNN with this fine tuning strategy (columns CNNf15
). From the table results it can be seen that CNNf15 is, most of the
times, better than the \textit{SRP-PHAT} baseline (except in two cases
for \seq03 in which there was a slight degradation).  The average
performance shows a consistent improvement of CNNf15 compared with
SRP-PHAT, between $1.8\%$ and $11.3\%$. However CNNf15 is still behind
GMBF in all cases but one (for \seq02 and $80 ms$).

\begin{table}[H]
\footnotesize
	\centering
	\begin{tabular}{c|cc|cc|cc}
    &   \multicolumn{2}{c|}{\textbf{80$ms$}}	& \multicolumn{2}{c|}{\textbf{160$ms$}} & \multicolumn{2}{c}{\textbf{320$ms$}}\\
    & GMBF & CNNf15 & GMBF & CNNf15 & GMBF & CNNf15 \\\hline\hline
    \multirow{2}{*}{\seq01 \legendMOTPDelta} & $\mbf{0.795}$ & $0.875$ & $\mbf{0.686}$ & $0.833$ & $\mbf{0.588}$ & $0.777$ \\
    & $\mbf{22.1\%}$ & $14.2\%$ & $\mbf{24.6\%}$ & $8.5\%$ & $\mbf{29.1\%}$ & $6.4\%$ \\\hline
    \multirow{2}{*}{\seq02 \legendMOTPDelta} & $0.864$ & $\mbf{0.839}$ & $\mbf{0.759}$ & $0.801$ & $\mbf{0.694}$ & $0.731$ \\
    & $10.0\%$ & $\mbf{12.6\%}$ & $\mbf{9.6\%}$ & $4.6\%$ & $\mbf{9.9\%}$ & $5.1\%$ \\\hline
    \multirow{2}{*}{\seq03 \legendMOTPDelta} & $\mbf{0.686}$ & $0.835$ & $\mbf{0.563}$ & $0.806$ & $\mbf{0.484}$ & $0.734$ \\
    & $\mbf{23.8}\%$ & $7.2\%$ & $\mbf{26.9\%}$ & -$4.7\%$ & $\mbf{29.9\%}$ & -$6.4\%$ \\\hline\hline
    \multirow{2}{*}{Average \legendMOTPDelta} & $\mbf{0.778}$ & $0.849$ & $\mbf{0.666}$ & $0.813$ & $\mbf{0.585}$ & $0.746$ \\
    & $\mbf{18.7\%}$ & $11.3\%$ & $\mbf{20.4\%}$ & $2.8\%$ & $\mbf{22.9\%}$ & $1.8\%$ \\
	\end{tabular}
	\caption{Results for the stratgy in \cite{velasco2012-F} (columns
    GMBF); and the CNN fine tuned with sequence \seq15
    (columns CNNf15).
}
	\label{tab:baselineResults+ft15}
\end{table}



Our conclusion is that the fine tuning procedure is able to effectively
complement the trained models from synthetic data, leading to results
that outperform SRP-PHAT. This is specially relevant as:

\begin{itemize}
\item The amount of fine tuning data is limited (only 36 seconds,
  corresponding to 436 frames, as shown in Table~\ref{tab:DataInfo}),
  thus opening the path to further improvements with a limited data
  recording effort.
\item The speaker used for fine tuning was mostly moving while speaking,
  while in the testing sequences the speakers are static while
  speaking. This means that the fine tuning material include far more
  active positions than in the testing sequences, and the network is
  able to extract the relevant information for the tested positions.
\item The speaker used for fine tuning is a male, and the obtained
  results for male speakers (sequences \seq01 and \seq03) and the female
  one (sequence \seq02) do not seem to show any gender-dependent bias,
  which means that the gender issue does not seem to play a role in the
  adequate adaptation of the network models.
\end{itemize}

When comparing the results of Table~\ref{tab:baselineResults} and
Table~\ref{tab:baselineResults+ft15}, and given the large improvement
when applying the fine tuning strategy, we could think that the effect
of the initial training with semi-synthetic data is limited. From this
argument, we run an additional training experiment in which we just
trained the network \emph{from scratch} using \seq15, aiming at
assessing the actual effect of semi-synthetic training+fine tuning versus
just training with real room data.

Table~\ref{tab:baselineResults+ft15vstr15} shows the comparison between
these two options: training from scratch using \seq15 (columns CNNt15)
and semi-synthetic training+fine tuning with \seq15 (columns CNNf15). The
average improvement of the latter approach varies between $1.8\%$ and
$11.3\%$ with an average improvement over all window lengths of $5.3\%$,
while the training from scratch average improvement varies between $-20.6\%$ and
$4.3\%$ with an average value of $-7.0\%$. These differences show that
the training+fine tuning proposal outperforms training the network from scratch, thus validating our methodology.

\begin{table}[H]
\footnotesize
	\centering
	\begin{tabular}{c|cc|cc|cc}
    &   \multicolumn{2}{c|}{\textbf{80$ms$}}	& \multicolumn{2}{c|}{\textbf{160$ms$}} & \multicolumn{2}{c}{\textbf{320$ms$}}\\
    & CNNt15 & CNNf15 & CNNt15 & CNNf15 & CNNt15 & CNNf15 \\\hline\hline
    \multirow{2}{*}{\seq01 \legendMOTPDelta} & $1.009$ & $\mbf{0.875}$ & $0.949$ & $\mbf{0.833}$ & $1.0009$ & $\mbf{0.777}$ \\
    & $1.1\%$ & $\mbf{14.2\%}$ & $-4.3\%$ & $\mbf{8.5\%}$ & $-21.6\%$ & $\mbf{6.4\%}$ \\\hline
    \multirow{2}{*}{\seq02 \legendMOTPDelta} & $\mbf{0.807}$ & $0.839$ & $\mbf{0.767}$ & $0.801$ & $0.807$ & $\mbf{0.731}$ \\
    & $\mbf{15.9\%}$ & $12.6\%$ & $\mbf{8.7\%}$ & $4.6\%$ & $-4.8\%$ & $\mbf{5.1\%}$ \\\hline
    \multirow{2}{*}{\seq03 \legendMOTPDelta} & $0.935$ & $\mbf{0.835}$ & $0.911$ & $\mbf{0.806}$ & $0.936$ & $\mbf{0.734}$ \\
    & $-3.9\%$ & $\mbf{7.2\%}$ & $-18.3\%$ & $\mbf{-4.7\%}$ & $-35.7\%$ & $\mbf{-6.4\%}$ \\\hline\hline
    \multirow{2}{*}{Average \legendMOTPDelta} & $0.915$ & $\mbf{0.849}$ & $0.875$ & $\mbf{0.813}$ & $0.916$ & $\mbf{0.746}$ \\
    & $4.3\%$ & $\mbf{11.3\%}$ & $-4.6\%$ & $\mbf{2.8\%}$ & $-20.6\%$ & $\mbf{1.8\%}$ \\
	\end{tabular}
	\caption{Results for the CNN proposal, either trained from scratch
    with sequence \seq15 (columns CNNt15) or fine tuned with sequence \seq15
    (columns CNNf15).
}
	\label{tab:baselineResults+ft15vstr15}
\end{table}



In spite of the relevant improvements with the fine tuning approach,
they are still far from making this suitable for further competitive
exploitation in the ASL scenario (provided we have the GMBF
alternative), so that we next aim at increasing the amount of fine
tuning material.



In our third experiment, we applied the fine tuning procedure using an
additional \emph{moving speaker} sequence, that is, including
\seq15 and \seq11 in the fine tuning subset.


Table~\ref{tab:baselineResults+ft15+11} shows the results obtained by
GMBF and CNN fine tuned with \seq15 and \seq11 (CNNf15+11 columns). In
this case, we see additional improvements over using only \seq15 for
fine tuning, and there is only one case in which CNNf15+11 does not
outperforms SRP-PHAT (with a marginal degradation of $-0.3\%$).

\begin{table}[H]
  \footnotesize
	\centering
	\begin{tabular}{c|cc|cc|cc}
    &   \multicolumn{2}{c|}{\textbf{80$ms$}}	& \multicolumn{2}{c|}{\textbf{160$ms$}} & \multicolumn{2}{c}{\textbf{320$ms$}}\\
    & GMBF & CNNf15+11 & GMBF & CNNf15+11 & GMBF & CNNf15+11 \\\hline\hline
    \multirow{2}{*}{\seq01 \legendMOTPDelta} & $\mbf{0.795}$ & $0.805$ & $\mbf{0.686}$ & $0.750$ & $\mbf{0.588}$ & $0.706$\\
    & $\mbf{22.1\%}$ & $21.1\%$ & $\mbf{24.6\%}$ & $17.6\%$ & $\mbf{29.1\%}$ & $14.9\%$\\\hline
    \multirow{2}{*}{\seq02 \legendMOTPDelta} & $0.864$ & $\mbf{0.809}$ & $0.759$ & $\mbf{0.716}$ & $\mbf{0.694}$ & $0.712$\\
    & $10.0\%$ & $\mbf{15.7\%}$ & $9.6\%$ & $\mbf{14.8\%}$ & $9.9\%$ & $7.5\%$\\\hline
    \multirow{2}{*}{\seq03 \legendMOTPDelta} & $\mbf{0.686}$ & $0.792$ & $\mbf{0.563}$ & $0.732$ & $\mbf{0.484}$ & $0.692$\\
    & $\mbf{23.8\%}$ & $12.0\%$ & $\mbf{26.9\%}$ & $4.9\%$ & $\mbf{29.9\%}$ & $-0.3\%$\\\hline\hline
    \multirow{2}{*}{Average \legendMOTPDelta} & $\mbf{0.778}$ & $0.802$ & $\mbf{0.666}$ & $0.732$ & $\mbf{0.585}$ & $0.703$\\
    & $\mbf{18.7\%}$ & $16.2\%$ & $\mbf{20.4\%}$ & $12.4\%$ & $\mbf{22.9\%}$ & $7.5\%$\\
	\end{tabular}
	\caption{Relative improvements over SRP-PHAT for the strategy in \cite{velasco2012-F} (columns
    GMBF); and the CNN fine tuned with sequences
    \seq15 and \seq11 (columns CNNf15+11) 
  }
	\label{tab:baselineResults+ft15+11}
\end{table}

The CNN based approach shows again an average consistent improvement
compared with SRP-PHAT between $7.5\%$ and $16.2\%$. 

In this case, the newly added sequence (\seq11, with a duration of only
33 seconds) for fine tuning corresponds to a randomly moving male
speaker, and the results show that its addition contributes to further
improvements in the CNN based proposal, but it is still behind GMBF in
all cases but two, but with results getting closer. This suggests that a
further increment in the fine tuning material should be considered.

Our last experiment will consist of fine tuning the network including
also additional static speaker sequences. To assure that the training
(including fine tuning) and testing material are fully independent, we
will fine tune with \seq15, \seq11 and with the static sequences that
are not tested in each experiment run, as shown in Table~\ref{tab:fine-tuning-material}.

\begin{table}[H]
	\centering
	\begin{tabular}{cc}
		\hline
		\textbf{Test sequence}	& \textbf{Fine tuning sequences} \\\hline\hline
		seq01 & \seq15 + \seq11 + \seq02 + \seq03 \\\hline
		seq02 & \seq15 + \seq11 + \seq01 + \seq03 \\\hline
		seq03 & \seq15 + \seq11 + \seq01 + \seq02 \\\hline
	\end{tabular}
	\caption{Fine tuning material used in the experiment corresponding to Table~\ref{tab:baselineResults+ft15+11+1+2+3} columns CNNf15+11+st.}
	\label{tab:fine-tuning-material}
\end{table}

Table~\ref{tab:baselineResults+ft15+11+1+2+3} shows the results obtained
for this fine tuning scenario, and the main conclusions are:

\begin{itemize}
\item The CNN based method exhibits much better average behavior than
  GMBF for all window sizes. 
Average absolute improvement against
  SRP-PHAT for the CNN is more than 10 points higher than for GMBF, reaching
  $31.3\%$ in the CNN case and $20.7\%$ for GMBF.
\item Considering individual sequences, CNN is significantly better
  than GMBF for sequences \seq01 and \seq02, and
  slightly worse for  \seq03.
\item Considering the best individual result, maximum improvement for
  the CNN is $41.6\%$ (\seq01, $320 ms$), while the top result for GMBF
  is $29.9\%$ (\seq03, $320 ms$).
\item The effect of adding static sequences is beneficial, as expected,
  provided that the acoustic tuning examples will be generated from
  positions which are similar, but not identical, as the speakers have
  varying heights and their position in the room is not strictly equal
  from sequence to sequence.
\item The improvements obtained are significant and come at the cost of
  additional fine tuning sequences. However, this extra cost is still
  reasonable, as the extra fine tuning material is of limited duration,
  around $400$ seconds in average ($6.65$ minutes).
\end{itemize}

\begin{table}[H]
  \footnotesize
	\centering
	\begin{tabular}{c|cc|cc|cc}
    &   \multicolumn{2}{c|}{\textbf{80$ms$}}	& \multicolumn{2}{c|}{\textbf{160$ms$}} & \multicolumn{2}{c}{\textbf{320$ms$}}\\
    & GMBF & CNNf15+11+st & GMBF & CNNf15+11+st & GMBF & CNNf15+11+st \\\hline\hline
    \multirow{2}{*}{\seq01 \legendMOTPDelta} & $0.795$ & $\mbf{0.607}$ & $0.686$ & $\mbf{0.540}$ & $0.588$ & $\mbf{0.485}$\\
    & $22.1\%$ & $\mbf{40.5\%}$ & $24.6\%$ & $\mbf{40.7\%}$ & $29.1\%$ & $\mbf{41.6\%}$ \\\hline
    \multirow{2}{*}{\seq02 \legendMOTPDelta} & $0.864$ & $\mbf{0.669}$ & $0.759$ & $\mbf{0.579}$ & $0.694$ & $\mbf{0.545}$ \\
    & $10.0\%$ & $\mbf{30.3\%}$ & $9.6\%$ & $\mbf{31.1\%}$ & $9.9\%$ & $\mbf{29.2\%}$ \\\hline
    \multirow{2}{*}{\seq03 \legendMOTPDelta} & $\mbf{0.686}$ & $0.707$ & $\mbf{0.563}$ & $0.617$ & $\mbf{0.484}$ & $0.501$ \\
    & $\mbf{23.8\%}$ & $21.4\%$ & $\mbf{26.9\%}$ & $19.9\%$ & $\mbf{29.9\%}$ & $27.4\%$ \\\hline\hline
    \multirow{2}{*}{Average \legendMOTPDelta} & $0.778$ & $\mbf{0.664}$ & $0.666$ & $\mbf{0.581}$ & $0.585$ & $\mbf{0.511}$ \\
    & $18.7\%$ & $\mbf{30.6\%}$ & $20.4\%$ & $\mbf{30.6\%}$ & $22.9\%$ & $\mbf{32.8\%}$ 
	\end{tabular}
	\caption{ Relative improvements over SRP-PHAT for the strategy in \cite{velasco2012-F} (columns
    GMBF); and the CNN fine tuned with the sequences
    described in Table~\ref{tab:fine-tuning-material} (columns CNNf15+11+st) 
  }
	\label{tab:baselineResults+ft15+11+1+2+3}
\end{table}

\begin{figure}[H]
  \centering
  \includegraphics[width=0.9\textwidth]{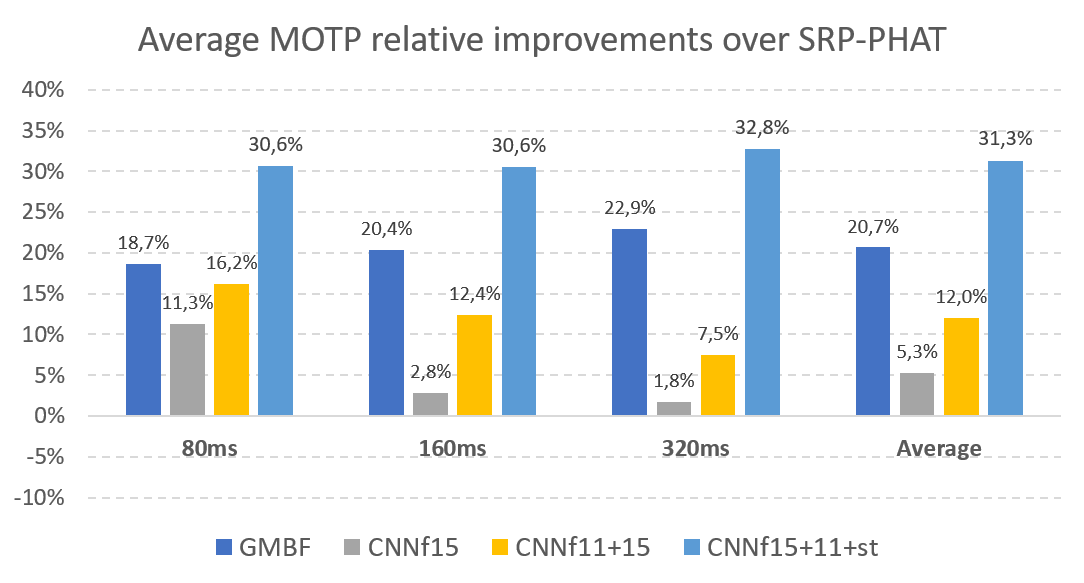}
	\caption{MOTP relative improvements over SRP-PHAT for GMBF and CNN
    using different fine tuning subsets (for all window sizes).}
	\label{fig:finalComparison80160320}
\end{figure}

Finally, to summarize, Figure~\ref{fig:finalComparison80160320} shows the
average MOTP relative improvements over \textit{SRP-PHAT} obtained by our CNN proposal using
different fine tuning subsets, and its comparison with the GMBF results,
for all the signal window sizes.




From the results obtained by our proposal, it is clear that the highest
contribution to the improvements from the bare CNN training is the fine
tuning procedure with limited data (CNNf15, comparing
Tables~\ref{tab:baselineResults} and~\ref{tab:baselineResults+ft15}),
while the addition of additional fine tuning material consistently
improves the results (Tables~\ref{tab:baselineResults+ft15+11}, and~\ref{tab:baselineResults+ft15+11+1+2+3}). It is again worth noticing that these improvements
are consistently independent of the gender of the considered speaker and
whether there is a match or not between the static or dynamic activity
of the speakers being used in the fine tuning subsets. This suggest that
the network is actually learning the acoustic cues that are related to
the localization problem, so that we can conclude that our proposal is a
suitable and promising strategy for solving the ASL task.



\section{Conclusions}
\label{sec:conclusions}


We have presented in this paper the first audio localization CNN that is
trained end-to-end from the audio signals to the source position. We
show that this method is very promising, outperforming the
state-of-the-art methods \cite{velasco2012-F,JoseVelascoThesis2017} and
those using \textit{SRP-PHAT}, given that sufficient fine tuning data is
available. In addition, our experiments show that the CNN method
exhibits good resistance against varying gender of the speaker and
different window sizes compared with the baseline methods.  Given that
the amount of data recordings for audio localization is limited at the
moment, we have thus proposed in the paper to first train the network
using semi-synthetic data followed by fine tuning using a small amount
of real data. This has been a common strategy in other fields to prevent
overfitting, and we show in the paper that it significantly improves the
system performance as compared with training the network from scratch
using real data.

In a future line of work we plan to improve the generation of
semi-synthetic data including reverberation effects and testing in
detail the effects of gender and language in the system performance. In
addition we plan to include more real data by developing a large corpus
for audio localization, that will be made available to the scientific
community for research purposes. Also, an extensive evaluation will be
carried out to asses the impact of the proposal with more complex
acquisition scenarios (comprising a higher number of microphone pairs).




\vspace{6pt} 



\authorcontributions{Conceptualization, Daniel Pizarro; Methodology,
  Writing - review \& editing and visualization, Daniel Pizarro, Juan Manuel Vera-Diaz and Javier Macias-Guarasa;
  Investigation, Juan Manuel Vera-Diaz; Writing - original draft, Juan
  Manuel Vera-Diaz; Software, Daniel Pizarro and Juan Manuel Vera-Diaz;
  Resources Javier Macias-Guarasa; Funding Acquisition, Daniel Pizarro
  and Javier Macias-Guarasa}

\funding{Parts of this work were funded by the Spanish Ministry of
  Economy and Competitiveness under projects HEIMDAL
  (TIN2016-75982-C2-1-R), ARTEMISA (TIN2016-80939-R), and SPACES-UAH
  (TIN2013-47630-C2-1-R), and by the University of Alcal\'a under
  projects CCGP2017/EXP-025 and CCG2016/EXP-010. Juan Manuel Vera-Diaz
  is funded by Comunidad de Madrid and FEDER under contract reference
  number
  PEJD-2017-PRE/TIC-4626.}


\conflictsofinterest{The authors declare no conflict of interest.}

\externalbibliography{yes}
\bibliography{paper}



\end{document}